# Robust static output feedback control using a PSO-DE/LMI hybrid algorithm

IlGon Ho, YongHyok Ri

Kim Il Sung University, Pyongyang

Democratic People's Republic of Korea

**Abstract**

Under the multi-objective framework, this paper presents a hybrid algorithm to solve robust static output feedback control problem for continuous polytopic uncertain system. To obtain static output feedback gain, a new hybrid algorithm is proposed by combination of a hybrid algorithm of the Particle Swarm Optimization (PSO) and Differential Evolution (DE), and the linear matrix inequality (LMI) method. The proposed algorithm is used to solve a optimization problem with a bilinear matrix inequality (BMI) constraints. The PSO-DE hybrid algorithm was used to obtain a population of controllers, and LMI approach was used to optimize a performance criterion of the system. Taking a hybrid $H_2/H_\infty$ control problem as example, the detailed algorithm is presented to solve robust static output feedback control problem. The simulation results show that the proposed hybrid algorithm improves the convergence rate and accuracy of iterative methods and the original DE-LMI algorithm.

**Key words**：static output feedback; particle swarm optimization algorithm; differential evolution algorithm; linear matrix inequality;

**Introduction**

Static output feedback controller is performed by constant matrix, so its algorithm is easy to realize and has no strong request for hardware.

With regard to realization of control systems, it is very important to design the multi-objective robust static output feedback controller. However, these problems are usually came to optimization problem with bilinear matrix inequalities(BMI) constraints[1]. BMI problem is NP-hard as non-convex optimization problem. Therefore it is impossible to solve with only LMI optimization. This motivates to develop the iterative LMI(ILMI) algorithm[2], which has calculation complexity and some limitation of application. And the method based on differential evolution(DE) algorithm and LMI is proposed[10]. Because DE algorithm has the conservation capability of the variety and fine search capability, is widely used to complex global optimization problems. Shortcomings is that rate of convergence is slow in late step and has early convergence.

Particle swarm optimization(PSO) algorithm, which is proposed by Kennedy and Eberhart is also optimal search mode based on swarm evolution. This method has noncomplex modeling, simple reality, fine rate of convergence and so on. However, PSO has also early convergence. PSO algorithm is widely used to optimization of objective function with no differentiability, discontinuity, multi-modal and so on [6]. Many researches improved the performance and convergence of the PSO by developing different variety and hybrid algorithm[7-8].

For solving of the BMI optimization problem we proposed a hybrid algorithm(PSO-DE/LMI) based on LMI optimization and design the robust multi-objective static output feedback controller for continuous polytopic uncertain system using it.

**1. Control problem formulation**

In realization of control systems the performance index for closed loop is all-round. For example, fine disturbance rejection, robust stability and dynamic performance are requested in the same time. Because this problem cannot be solved by design with regard to single objective, design problem of the multi-objective controller is very important in practice.

Polytopic uncertain system is described as state space model as follows;

$$\begin{cases} \dot{x} = Ax + B_1 w + B_2 u \\ z_1 = C_1 x + D_{11} w + D_{12} u \\ z_2 = C_2 x + D_{21} w + D_{22} u \\ y = Cx \end{cases} \quad (1)$$

Here $x \in \mathbf{R}^n$ is the state, $u \in \mathbf{R}^m$ is the control input, $w \in \mathbf{R}^l$ is the disturbance input, $y \in \mathbf{R}^p$ is the measured output, $z_1 \in \mathbf{R}^{p_1}$ and $z_2 \in \mathbf{R}^{p_2}$ are evaluation outputs. Uncertain system matrices is described by the below known vertex matrices as follows;

$$\begin{bmatrix} A & B_1 & B_2 \\ C_1 & D_{11} & D_{12} \\ C_2 & D_{21} & D_{22} \end{bmatrix} = \sum_{i=1}^{N} \alpha_i \begin{bmatrix} A_i & B_{1i} & B_{2i} \\ C_{1i} & D_{11i} & D_{12i} \\ C_{2i} & D_{21i} & D_{22i} \end{bmatrix}, \quad \alpha_i \in \Omega,$$

$$\Omega = \left\{ \alpha \in \mathbf{R}^N : \sum_{i=1}^{N} \alpha_i = 1, \alpha_i \geq 0, i = 1, \cdots, N \right\} \quad (2)$$

For representing the design method of the multi-objective robust static output feedback controller of this uncertain system, we consider the mixed $H_2/H_\infty$ control problem as an example. At this time $z_1$ and $z_\infty$ are evaluation outputs of the $H_\infty$ performance and $H_2$ performance, respectively.

For the system represented by equation (1), we consider the static output feedback by constant matrix $K$ as follows;

$$u = Ky, \quad K \in \Gamma \quad (3)$$

Then closed-loop is

$$\begin{cases} \dot{x} = (A + B_2 KC)x + B_1 w \\ z_\infty = (C_1 + D_{12} KC)x + D_{11} w \\ z_2 = (C_2 + D_{22} KC)x + D_{21} w \end{cases} \quad (4)$$

Here $\Gamma$ is bounded matrix set, which stabilize the closed-loop asymptotically. Structure of the feedback control system is shown as Fig 1.

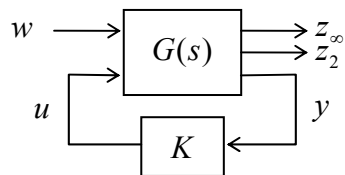

Fig.1. Structure of the feedback control system

For the closed system (4) transfer function(matrix) from $w$ to $z_\infty$ is given by $G_{z_\infty w}(s)$ and transfer function(matrix) from $w$ to $z_2$ is given by $G_{z_2 w}(s)$. We consider mixed $H_2/H_\infty$ control problem is formulated as follows;

The problem is to decide the static output feedback controller $K \in \Gamma$, which minimize the $H_2$ performance from $w$ to $z_2$ with stabilizing the closed-loop. This is equal to existence of symmetric

positive define matrix $X_2$, $X_\infty$ satisfying the LMIs given by[9]:

$$\min_{K \in \Gamma} \delta \quad \text{s.t.}$$

$$\begin{bmatrix} (A_i + B_{2i}KC)^T X_2 + X_2(A_i + B_{2i}KC) & X_2 B_{1i} \\ * & -I \end{bmatrix} < 0,$$

$$\begin{bmatrix} X_2 & (C_{1i} + D_{12}KC)^T \\ * & S \end{bmatrix} > 0, \quad \text{Trace}(S) < \delta,$$

$$\begin{bmatrix} (A_i + B_{2i}KC)^T X_\infty + X_\infty(A_i + B_{2i}KC) & * & * \\ B_{1i}^T X_\infty & -\gamma I & * \\ (C_{1i} + D_{12i}KC) & 0 & -\gamma I \end{bmatrix} < 0,$$

$$X_\infty > 0, \quad X_2 > 0, \quad i = 1, \cdots, N \quad (5)$$

Here $\Gamma := \{K \mid \|G_{z_2 w}\|_2^2 \leq \delta\}$, $I$ and $\mathbf{0}$ are unit matrix and zero matrix, respectively, "*" denotes transposed matrix for diagonal opposite of appropriate diagonal matrix.

When we design the mixed multi-objective $H_2/H_\infty$ controller, promise is assumption of $X_2 = X_\infty$. But this assumption often leads to additional conservation of controllers. In this paper we design the controller without this assumption to decrease such conservation.

For each vertex(for example, in case $i=1$) model, all of the inequality constraints given by (5) are nonlinear with regard to unknown matrix variable $K$, $X_2$ and $X_\infty$. In other words, all of these constraints become to bilinear matrix inequalities(BMI), so the problem given by (5) come to optimization problem with BMI constraints. As you know, BMI problems are non-convex and NP-hard. It is difficult to solve these problems only with LMI or Riccati inequality. Therefore, we solve the BMI optimization problem as above, by associating the PSO-DE hybrid algorithm and LMI optimization. That is, controller $K$ is evolved by PSO-DE algorithm, in this time all of the inequality constraints given by (5) come to convex optimization. Therefore, approximate solution of the whole optimization problem is decided by solving the LMI optimization respect to matrix $X_2$, $X_\infty$.

**2. Design of the multi-objective static output feedback controller**

In this paper we decide the feedback gain by using the PSO-DE/LMI hybrid algorithm, which is constructed of combining the PSO-DE algorithm and LMI optimization. In the algorithm, number of the swarm's particles is $NP$, each particle is vector with dimension $D$. Then position of the $i$th particle of swarm is $X_i=(x_{i1}, x_{i2},\ldots, x_{iD})$ and velocity is $V_i=(v_{i1}, v_{i2},\ldots, v_{iD})$. Individual optimal position of the $i$th particle(i.e. recognition experience) is $P_i=(p_{i1}, p_{i2},\ldots, p_{iD})$ and optimal position of the swarm(i.e. population experience) is $P_{best}=(p_{best1}, p_{best2},\ldots, p_{bestD})$

Character of the PSO algorithm is that the optimization is achieved by sharing information between each particle(individual). That is, in the swarm, motion of each particle is decided by the recognition experience(experience in itself) and population experience(experience of the whole swarm). In PSO algorithm using compressibility factor, behavior of each particle is as follows[5].

$$V_i(t+1) = \chi(V_i(t) + c_1 r_1(P_i(t) - X_i(t)) + c_2 r_2(P_{best}(t) - X_i(t))) \quad (6)$$

$$X_i(t+1) = X_i(t) + V_i(t+1) \quad (7)$$

$$i = 1,2,\cdots, NP$$

Here $\chi$ is compressibility factor, $c_1$ and $c_2$ are recognition experience factor and population experience factor, are not negative, respectively. And $r_1$, $r_2$ are random numbers distributed on [0, 1]. Recognition experience factor and population experience factor control the influence of recognition experience of particle $P_i$ and population experience $P_{best}$, respectively. To guarantee fast rate of convergence, we can select as $\chi = 0.72984$, $c_1=c_2=2.05$[5]. In this paper we use the PSO-DE hybrid algorithm based on evolution of the recognition experience. Character of this algorithm is that it has a little imposition of computation and fine convergence than other PSO-DE algorithm[8].

In PSO-DE/LMI hybrid algorithm controller $K_i$ is corresponded to position of $i$th particle, that is, $X_i$ in (6) and (7). Let the set of the recognition experiences $S_p=\{P_1, P_2,…,P_{NP}\}$. After performing a round we apply a round DE algorithm to individual with change recognition experience(improvement). In other words, we apply three operation(mutation, cross, selection) of the DE algorithm only to recognition experience of "desirable" individual. Here so called "desirable" individual is meaning of the individual, have updated recognition experience from PSO. This method is very efficient than method, which evolve all individual of the swarm.

Flow chart of the PSO-DE/LMI algorithm is shown as fig 2.

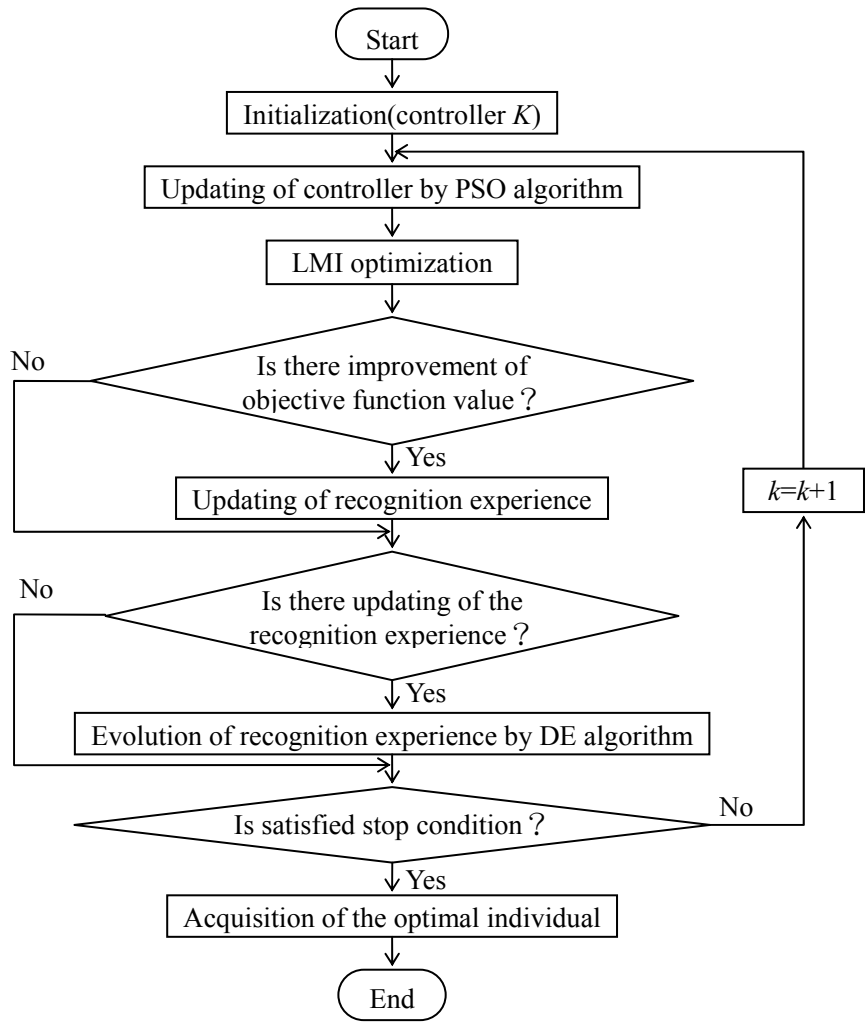

Fig.2 PSO-DE/LMI hybrid algorithm

Concrete step of this algorithm is as follows.
1: Each particle(controller $K_i(0)$, $i=1,…,NP$) of swarm is initialized in given region. If $K_i(0)$ doesn't satisfy the LMI constraint (5)(in other words, the LMI with regard to unknown matrices $X_2$ and $X_\infty$ is not solvable) we iterate the initialization until the constraint is satisfied. We also initialize each recognition experience $P_i(t)$. Size of swarm $NP$ is can be set to 5~10 times of number of controllers.
2: We get objective function value(fitness value) $f(K_i(0))$ by solving the LMI optimization with $X_2$ and $X_\infty$ for each given particle $K_i(0)$, $i=1,\ldots,NP$. We initialize the population experience $P_{best}(t)$ by using given fitness values.
3: **for** time step $t$ **do**
4:   **for** particle number $i$ **do**
5:     We update $K_i(t)$ by (6),(7). If updated $K_i(t)$ does not satisfy LMI constraint (5), we iterate the updating until the constraint is satisfied.
6:     For given $K_i(t)$ we get fitness value $f(K_i(t))$ by solving LMI optimization with $X_2$ and $X_\infty$.
7:     If $f(K_i(t)) < f(P_i(t))$ then we update recognition experience $P_i(t)$. If also $f(P_i(t)) < f(P_{best}(t))$ then population experience $P_{best}(t)$.
8:     **If** $P_i(t)$ is changed **then**
9:       /* We evolve $P_i(t)$ by applying a round DE algorithm is set $S_p$. */
10:       By applying variation of DE in set $S_p$ we generate variation vector $v_t^i$.(variation operation)
11:       We generate testing vector $u_t^i$ by crossing variation vector $v_t^i$ and objective vector $P_i(t)$.(cross operation)
12:       If testing vector $u_t^i$ satisfy LMI constraint we get fitness value $f(u_t^i)$ by solving LMI optimization with $X_2$ and $X_\infty$. If the constraint is not satisfied we iterate the variation and cross operation until that is satisfied.
13:       If $f(u_t^i) < f(P_i(t))$ then we update particle's experience $P_i(t)$. If also $f(P_i(t)) < f(P_{best}(t))$ then we update population experience $P_{best}(t)$.(selection operation)
14:     **end if**
15:   **end for**
16: **end for**

## 3. Computational examples

**Example 1:** Consider the following system[10].

$$A = \begin{bmatrix} 0 & 1 \\ -1 & 0 \end{bmatrix}, \quad B_1 = I_2, \quad B_2 = \begin{bmatrix} 0 \\ 1 \end{bmatrix}, \quad C_1 = \mathbf{0}, \quad D_{11} = \mathbf{0}, \quad D_{12} = \mathbf{0}$$

$$C_2 = \begin{bmatrix} 1 & 0 \\ 0 & 0 \end{bmatrix}, \quad D_{21} = \mathbf{0}, \quad D_{22} = \begin{bmatrix} 0 \\ 1 \end{bmatrix}$$

In literature [10] optimal feedback gain obtained by optimal $H_2$ static output feedback control is $K=0.8198$ and minimum performance index(square of $H_2$ norm) is 2.4495.

In proposed PSO-DE/LMI hybrid algorithm, we obtain the optimal $H_2$ performance 2.4495 when we set size of swarm and generation to 5 and 5. For comparison we design the controller by using the DE-LMI algorithm proposed in literature [3] under same parameters and constraints as in PSO-DE/LMI algorithm.

For two algorithms relation between number of iterations and performance index(objective function values) is shown as fig 3. From fig we know that optimal $H_2$ performance is achieved only with over 5 generations in PSO-DE/LMI algorithm. However that is achieved with over 11 generations in DE-LMI algorithm. Of course at each generation computational amount of PSO-DE/LMI algorithm is more than DE-LMI a little, but is not more than DE, because their difference is computational amount of PSO(equation (6) and (7)). On the whole, the computational amount of PSO-DE/LMI hybrid algorithm for 5 generations is much less than one of the DE-LMI algorithm for 11 generations. This means that rate of convergence and accuracy of PSO-DE/LMI hybrid algorithm is predominant than DE-LMI algorithm.

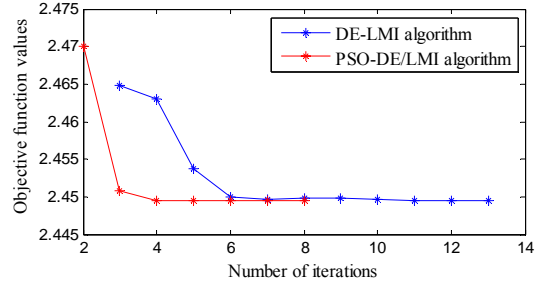

Fig.3 Relation between number of iterations and objective function values

**Example 2:** Consider the system, matrix $A$ has polytopic uncertainty as follows[10];

$$A = \begin{bmatrix} -2.9800 & a_{12} & 0 & -0.0340 \\ -0.9900 & -0.2100 & 0.0350 & -0.0011 \\ 0 & 0 & 0 & 1.0000 \\ 0.3900 & -5.5550 & 0 & -1.8900 \end{bmatrix}$$

$$B_2 = \begin{bmatrix} -0.0320 \\ 0 \\ 0 \\ -1.6000 \end{bmatrix}, \quad D_{22} = \begin{bmatrix} 1 \\ 0 \\ 0 \\ 0 \end{bmatrix}, \quad C = \begin{bmatrix} 0 & 0 \\ 0 & 0 \\ 1 & 0 \\ 0 & 1 \end{bmatrix}$$

$$B_1 = C_1 = C_2 = I_4, \quad D_{11} = \mathbf{0}, \quad D_{12} = \mathbf{0}, \quad D_{21} = \mathbf{0}$$

Here $a_{12} \in [-0.57 \quad 2.43]$.

When we have size of swarm 5 and 10 generations with $H_2$ performance constraint 20 for PSO-DE/LMI hybrid algorithm, static output feedback gain $K$=[2.2422　2.6117] and optimal performance index $\|H\|_2^2 = 94.9897$ are obtained. When we use the DE-LMI algorithm in literature [3], K=[2.0200　1.9552] and $\|H\|_2^2 = 110.8312$ are obtained under same conditions. This shows that the performance of the controller obtained by PSO-DE/LMI hybrid algorithm is predominant than literatures' results.

**Conclusion**

In this paper robust multi-objective static output feedback control is considered for the continuous systems with polytopic uncertainties. A hybrid algorithm, which is constructed by combining the PSO-DE algorithm and LMI optimization is proposed for solving of the BMI optimization, is non-convex and NP-hard.

Robustness, dynamic performance and convergence are improved by evolution of controller using PSO-DE hybrid algorithm based on evolution of recognition experience and optimization of the performance index using LMI. And proposed PSO-DE hybrid algorithm has more less conservation by excluding the assumption that matrix variables of the performance indices are equal

Also, we show improved control performance, rate of convergence and accuracy compa ring past results in simulations.

**References**

[1] Arzelier D, Peaucelle D. An iterative method for mixed $H_2/H_\infty$ synthesis via static output-feedback [C]. Proceedings of 41st IEEE Conference on Decision and Control. Las Vegas, Nevada, USA: IEEE, 2002:3464-3469

[2] Huang D, Nguang S K. Static output feedback controller design for fuzzy systems: An ILMI approach [J]. Information Sciences, 2007,177: 3005-3015

[3] Geromel J C, Peres P L D, Souza S R. Convex analysis of output feedback control problems: robust stability and performance [J]. IEEE Transaction on Automatic Control, 1996, 41(7): 997-1003

[4] Kennedy J, Eberhart R C. Particle Swarm Optimization [C]. Proceeding of the IEEE Service Center, 1995: 1941-1948.

[5] Clerc M, Kennedy J. The particle swarm - explosion, stability and convergence in a multidimensional complex space [J]. IEEE Transactions on Evolutionary Computation, 2002, 6(1) :58–73.

[6] Rana S, Jasola S. A review on particle swarm optimization algorithms and their applications to data clustering [J]. Artificial Intelligence Review, 2011, 35(3): 211-222.

[7] Zhang C, Ning J, Lu S, Ouyang D, and Ding T. A novel hybrid differential evolution and particle swarm optimization algorithm for unconstrained optimization [J]. Operations Research Letters, 2009, 37(2): 117-122.

[8] Epitropakis M G, Plagianakos V P, Vrahatis M N. Evolving cognitive and social experience in Particle Swarm Optimization through Differential Evolution [C]. 2010 IEEE Congress on Evolutionary Computation (CEC), 2010:1-8.

[9] Boyd S, et al. Linear Matrix Inequalities in System and Control Theory [M]. Philadelphia: *SIAM*, 1994.

[10] 刘树博. 基于新型优化算法的主动悬架鲁棒输出反馈控制研究 [D]. 长春: 吉林大学机械学院，2010